\documentclass[aps]{revtex4}
\usepackage{eurosym}
\usepackage{amsfonts}
\usepackage{amsmath}
\usepackage{amssymb,epsf}
\usepackage{color}
\usepackage{graphicx}
\usepackage{epstopdf}
\usepackage{float}
\usepackage{caption}
\usepackage{subfig}

\begin{document}

\title{Contraction of cold neutron star due to in the presence a quark core}
\author{B. Eslam Panah$^{1,2,3}$\footnote{%
email address: beslampanah@shirazu.ac.ir}, T. Yazdizadeh$^{4,5}$ \footnote{%
email address: tyazdizadeh@yahoo.com} and G. H. Bordbar $^{1,6}$ \footnote{%
email address: ghbordbar@shiraz.ac.ir} }
\affiliation{$^1$ Physics Department and Biruni Observatory, College of Sciences, Shiraz
University, Shiraz 71454, Iran\\
$^2$ Research Institute for Astronomy and Astrophysics of Maragha (RIAAM),
P.O. Box 55134-441, Maragha, Iran\\
$^3$ ICRANet, Piazza della Repubblica 10, I-65122 Pescara, Italy\\
$^4$ Department of Physics, Payame Noor University (PNU), P.O. Box
19395-3697, Tehran, Iran\\
$^5$ Islamic Azad University, Bafgh Branch 89751-43398, Bafgh, Iran,\\
$^6$ Department of Physics and Astronomy, University of Waterloo, 200
University Avenue West, Waterloo, Ontario N2L3G1, Canada}

\begin{abstract}
Motivated by importance of the existence of quark matter on structure of
neutron star. For this purpose, we use a suitable equation of state (EoS)
which include three different parts: i) a layer of hadronic matter, ii) a
mixed phase of quarks and hadrons, and, iii) a strange quark matter in the
core. For this system, in order to do more investigation of the EoS, we
evaluate energy, Le Chatelier's principle and stability conditions. Our
results show that the EoS satisfies these conditions. Considering this EoS,
we study the effect of quark matter on the structure of neutron stars such
as maximum mass and the corresponding radius, average density, compactness,
Kretschmann scalar, Schwarzschild radius, gravitational redshift and
dynamical stability. Also, considering the mentioned EoS in this paper, we
find that the maximum mass of hybrid stars is a little smaller than that of
the corresponding pure neutron star. Indeed the maximum mass of hybrid stars
can be quite close to the pure neutron stars. Our calculations about the
dynamical stability show that these stars are stable against the radial
adiabatic infinitesimal perturbations. In addition, our analyze indicates
that neutron stars are under a contraction due to the existence of quark
core.
\end{abstract}

\maketitle

\section{Introduction}

Neutron stars which are born in the aftermath of core-collapsing supernova
(SN) explosions, are a cosmic laboratory and the best environment for the
studying dense matter problems. It is notable that, in the center of neutron
star because of high densities, the matter is envisaged to have a transition
from hadronic matter to strange quark matter, see refs. \cite{Ivanenko,
Itoh, Fritzsch,Collins}, for more details. Also, Glendenning in ref. \cite%
{glendenning1992}, showed that proper construction of phase transition
between the hadron and quark, inside the neutron stars implies the
coexistence of nucleonic matter and quark matter over a finite range of the
pressure. Accordingly, a mixed hadron-quark phase exists in the neutron
star, so that its energy is lower than that of the quark matter and
nucleonic matter. Phase transition between quark matter core and hadronic
external layers (hadron-quark phase) in neutron star is\ an interesting
subject which has been investigated by many authors \cite%
{Phase1,Phase2,Phase3,Phase4,Phase5,Phase6,Phase7,Phase8,Phase9,Phase10,Phase11,Phase12,Phase13}%
. For example,\ the effects of quark-hadron matter in the center of neutron
stars have been studied in ref. \cite{Orsaria2013a,Orsaria2013b}, and the
obtained results have been shown that for $PSR~J1614-2230$, and $%
PSR~J0348+0432$ with masses about $2M_{\,\odot }$, may contain a region of
quark-hybrid matter in their center. Plumari et al. have investigated the
effects of a quark core inside neutron star by considering the quark-gluon
EoS in the framework of field correlator model \cite{Plumari}. They have
found an upper limit for the mass of neutron stars by adjusting some
parameters. This limit was in the range $M_{\max }\simeq 2M_{\,\odot }$. In
addition, Chen et al. have studied cold dense quark matter and hybrid stars
with a Dyson-Schwinger quark model and various choices of the quark-gluon
vertex \cite{Chen}. They have showed that hadron states have the maximum
mass lower than the pure nucleonic neutron stars, but higher than two solar
masses. Their results depended on parameters of EoS. Also, Lastowiecki et
al. have found that compact stars masses of about $2M_{\,\odot }$ such as $%
PSR~J1614-2230$ and $PSR~J3048+0432$ were compatible with the possible
existence of deconfined quark matter in their core \cite{Lastowiecki}.
Neutrino emissivity in the quark-hadron mixed phase of neutron stars have
been investigated in ref. \cite{Spinella}. According to importance of
existence of quark matter in the neutron stars \cite%
{NSQMI,NSQMII,NSQMIII,NSQMIV,NSQMV,NSQMVI,NSQMVII,NSQMVIII,NSQMIX,NSQMIXa,NSQMX,NSQMXa, NSQMXI,NSQMXII,NSQMXIII,NSQMXIV,NSQMXV,NSQMXVa,NSQMXVI,NSQMXVIII,NSQMXXIa,NSQMXXIb,NSQMXXII, NSQMXXIII,NSQMXXIV,NSQMXXV,NSQMXXVI}%
, we consider a neutron star to be composed of a hadronic matter layer, a
mixed phase of quarks and hadrons, and in the core of star, a quark matter.
One of our aims in this work is determining the structure of neutron star
with a quark core and comparing it with observation data.

In order to study the structure of stars and their phenomenological
properties, we must use the hydrostatic equilibrium equation (HEE). Indeed,
this equation is based on the fact that a typical star will be in
equilibrium when there is a balance between the gravitational force and the
internal pressure. The first HEE equation was introduced by Tolman,
Oppenheimer and Volkoff (TOV) \cite{Tolman1934, Tolman1939, Oppenheimer} in
the Einstein gravity which is known as TOV equation. Considering TOV
equation, the structure of compact stars have been evaluated by many authors
in refs. \cite{TOVI,TOVII,TOVIII,TOVIV,TOVV,TOVVII,TOVIX,TOVXII,TOVXIII}.

According to this fact that, there are some massive neutron stars with mass
more than two times of solar mass, $M\geq 2M_{\,\odot }$, for example, $%
4U1700-377$ with $M=2.4M_{\,\odot }$\ \cite{ClarK}, and $J1748-2021B$ with $%
M=2.7M_{\,\odot }$\ \cite{J1748-2021B}, one of our goals in this work is
related to answer this question: is there a quark core inside neutron stars
with mass more than $2M_{\,\odot }$\ ($M\geq 2M_{\,\odot }$), by considering
a suitable EoS which obtained by combining three different parts, a layer of
hadronic matter, a mixed phase of quarks and hadrons, and a strange quark
matter in core? In order to evaluate a suitable EoS, we will study energy,
Le Chatelier's principle and stability conditions of this EoS. Another our
goal is related to the investigation of the effects of quark core in the
structure of hybrid star. For this goal, we study the difference between the
structure of hybrid stars and pure neutron stars.

The outline of the paper is as follows; First, in order to investigate the
structure of neutron stars, we evaluate a suitable EoS which includes three
different layers. Then, we compare the structure of neutron stars by
considering the EoS with and without a quark core. Indeed, we study the
effects of EoS by applying a quark core on the structure of neutron stars.
Next, we compare our obtained results with those of observational data. The
last section is devoted to closing remarks.

\section{equation of state}

\label{Energy calculation} A neutron star with a quark core composed of a
hadronic matter layer, a mixed part of quarks and hadrons and a quark matter
in core. Thus we calculate the EoS of different parts of this star in the
following subsections.


\subsection{Hadron Phase}

We use the lowest order constrained variational (LOCV) many-body method to
determine the EoS of nucleonic matter \cite{bordbar1997, bordbar1998,
modarres1998,bigdeli2009}. We consider a cluster expansion of the energy
functional up to the two-body term,
\begin{equation}
E([f])=\frac{<\psi \mid H\mid \ \psi >}{N<\psi \mid \psi >}=E_{1}+E_{2},
\label{eq1}
\end{equation}%
in which $H$ is the Hamiltonian of the system. Also, $\psi $ is the total
wave function in which we consider a trail many-body wave function as $\psi
=F\phi $. Here $\phi $ is a uncorrelated ground-state wave function of $N$
independent nucleons, and $F$ is a proper $N$-body correlation function
which is taken according to the Jastrow ansatz, $F=\mathcal{S}%
\prod_{i>j}f(ij)$, in which $\mathcal{S}$ and $f(ij)$ are the symmetrizing
operator and the two-body correlation function, respectively. Using the
Jastrow ansatz and after some algebra, the energy is calculated (see \cite%
{clark}, for more details). The one body term is $E_{1}=\sum_{i=1,2}{\frac{3%
}{5}\frac{\hbar ^{2}k_{i}^{2}}{2m_{i}}\frac{\rho _{i}}{\rho }}$, for an
asymmetrical nucleonic matter, where\emph{\textbf{\ }}$\rho _{i}$ are the
nucleonic densities associated with the protons and neutrons $(\rho =\rho
_{p}+\rho _{n})$ and $k_{i}=(6\pi ^{2}\rho _{i})^{1/3}$ is the Fermi
momentum of particle $i$. The two-body energy is $E_{2}=\frac{1}{2N}%
\sum_{ij}<ij|\nu (12)|ij-ji>$. In this relation, the operator $\nu (12)$ is
the effective nuclear potential. The complete calculation for nuclear matter
has been presented in ref. \cite{bordbar1998}. It should be noted that for
calculation of the structure properties of neutron star, the equation of
state of hadronic matter has a crucial role. On the other hand, various
nucleon-nucleon potentials lead to the different equations of state.
Therefore, the maximum mass of neutron star depends highly on the
inter-nucleon potential \cite{bordbar2006}. Here the three-body
nucleon-nucleon potential is very important for the nucleonic matter
calculations. In the present paper, for the hadronic matter calculations, we
use the $UV14+TNI$ potential in which the effect of three-body nuclear force
has been considered.


\subsection{Quark Phase}

The total energy of strange quark matter with deconfined up (u), down (d)
and strange (s) quarks within MIT bag model \cite{chodos1974, baym1985} is
given by
\begin{equation}
\mathcal{E}_{tot}=\mathcal{E}_{u}+\mathcal{E}_{d}+\mathcal{E}_{s}+B.
\label{eq4}
\end{equation}
The quark confinement in MIT bag model is satisfied by a density dependent
bag constant $B$, that is interpreted as the difference between energy
densities of non interacting and interacting quarks. We use a density
dependent with the Gaussian form \cite{TOVVI,burgio2002},
\begin{equation}
B(\rho )=B_{\infty }+(B_{0}-B_{\infty })\exp \left[ {-\beta (\frac{\rho }{%
\rho _{0}})^{2}}\right] .  \label{eq5}
\end{equation}
In the equation (\ref{eq4}) $\mathcal{E}_{i}$ is
\begin{equation*}
\mathcal{E}_{i}=\frac{3m_{i}^{4}}{8\pi ^{2}}\left[ x_{i}(2x_{i}^{2}+1)(\sqrt{%
1+x_{i}^{2}})-\sinh ^{-1}x_{i}\right] ,
\end{equation*}%
where $x_{i}=\frac{k_{F}^{(i)}}{m_{i}}$ and, $k_{F}^{(i)}=(\rho _{i}\pi
^{2})^{1/3}$ is the fermi momentum of quark $i$. See ref. \cite{TOVVI} for
more details.

Now, by using the energy density from Eq. (\ref{eq4}), we can obtain the EoS
of quark matter in the MIT bag model,
\begin{equation}
P(\rho )=\rho \frac{\partial \mathcal{E}}{\partial \rho }-\mathcal{E}.
\label{eq7}
\end{equation}

\subsection{Mixed phase}

The hadron-quark phase happens in the high range of baryon density values.
The occupied fraction of space by quark matter smoothly increases from zero
where there is no quark to unity when the last nucleons dissolve into the
quarks. In this phase, there are a mixture of hadrons, quarks and electrons.
According to the Gibss equilibrium condition, the pressures, the
temperatures and chemical potentials of both hadron and quark phases are
equal \cite{glendenning1992}, $(\mu _{p}^{Q}=\mu _{p}^{H})$ and $(\mu
_{n}^{Q}=\mu _{n}^{H})$ where $\mu _{n}^{H}$ ($\mu _{p}^{H}$) and $\mu
_{n}^{Q}$ ($\mu _{p}^{Q}$) are neutrons (protons) chemical potential in the
hadron phase and the quark phase, respectively. It is notable that $\mu _{n}$
and $\mu _{p}$ are given by
\begin{equation}
\mu _{n}=\frac{\partial \mathcal{E}}{\partial \rho _{n}},~~~~~\mu _{p}=\frac{%
\partial \mathcal{E}}{\partial \rho _{p}}.  \label{eq10}
\end{equation}

As the chemical potentials determine the charge densities, the volume
fraction occupied by quark matter, $\chi $, can be obtained by the
requirement of global charge neutrality. Then the total energy density and
baryon density of mixed phase could be determined,
\begin{eqnarray}
\chi (\frac{2}{3}\rho _{u}-\frac{1}{3}\rho _{d}-\frac{1}{3}\rho
_{s})+(1-\chi )\rho _{p}-\rho _{e} &=&0,  \label{eq12} \\
\chi \rho _{Q}+(1-\chi )\rho _{H} &=&\rho _{B},  \label{eq13} \\
\chi \mathcal{E}_{QP}+(1-\chi )\mathcal{E}_{HP} &=&\mathcal{E}_{MP}.
\label{eq14}
\end{eqnarray}
Calculations regarding the EoS of mixed phase has been fully discussed in
Ref. \cite{TOVVI}.

At this stage we can determine EoS of neutron star with quark core using the
results of proceeding sections. Also we investigate the energy and stability
conditions for our results. For this purpose at first, we extract a
mathematical form for the EoS presented as a polynomial function in the
following form
\begin{equation}
P=\sum_{i=1}^{7}a_{i}\mathcal{E}^{7-i},  \label{Prho}
\end{equation}%
where $a_{i}$ are
\begin{eqnarray}
a_{1} &=&1.194\times 10^{-57},~\ \ a_{2}=-0.2467\times 10^{-40},  \notag \\
a_{3} &=&2.011\times 10^{-25},~~~a_{4}=-8.123\times 10^{-10},  \notag \\
a_{5} &=&1.656\times 10^{6},~~~\ \ \ a_{6}=-1.201\times 10^{21},  \notag \\
a_{7} &=&2.915\times 10^{35}.
\end{eqnarray}

In order to do more investigation of the obtained EoS, we evaluate energy, {%
Le Chatelier's principle} and stability conditions as follows.

\subsection{Energy conditions}

It is expected that the energy conditions are satisfied by any EoS of
matter. In general, there are four energy conditions. The null energy
condition (NEC), weak energy condition (WEC), strong energy condition (SEC)
and dominant energy condition (DEC). The two first cases state that for any
matter distribution, energy should be non-negative. The weak energy apply to
time-like vectors and the null energy condition is for null vectors. The
strange energy condition states that the gravitational field is attractive.
The last condition means that the energy flow rate of matter is less than
the speed of light \cite{Poisson}. These conditions in the center of neutron
star are given as follows
\begin{eqnarray}
NEC &\rightarrow &P_{c}+\mathcal{E}_{c}\geq 0,  \label{NEC} \\
WEC &\rightarrow &P_{c}+\mathcal{E}_{c}\geq 0\quad \&\quad \mathcal{E}%
_{c}\geq 0,  \label{WEC} \\
SEC &\rightarrow &P_{c}+\mathcal{E}_{c}\geq 0\quad \&\quad 3P_{c}+\mathcal{E}%
_{c}\geq 0,  \label{SEC} \\
DEC &\rightarrow &\mathcal{E}_{c}>|P_{c}|,  \label{DEC}
\end{eqnarray}%
where $\mathcal{E}_{c}$ and $P_{c}$ are the density and pressure in the
center of neutron star ($r=0$), respectively. The results of above
conditions for our EoS are given in Table \ref{tab1}.

\begin{table*}[tbp]
\caption{Energy conditions for neutron star with a quark core.}
\label{tab1}
\begin{center}
\begin{tabular}{||c|c|c|c|c|c|c||}
\hline\hline
Type of Star & $\mathcal{E}_c\left({10^{14}gr/cm^3}\right)$ & $P_c\left({%
10^{14}gr/cm^3}\right)$ & NEC & WEC & SEC & DEC \\ \hline\hline
$NS+Quark Core$ & 25.8 & 7.8 & $\surd$ & $\surd$ & $\surd$ & $\surd$ \\
\hline\hline
\end{tabular}%
\end{center}
\end{table*}
%
We see that our EoS satisfied all mentioned energy conditions.

\subsection{Stability}

\begin{figure}[tbp]
$%
\begin{array}{c}
\epsfxsize=10cm \epsffile{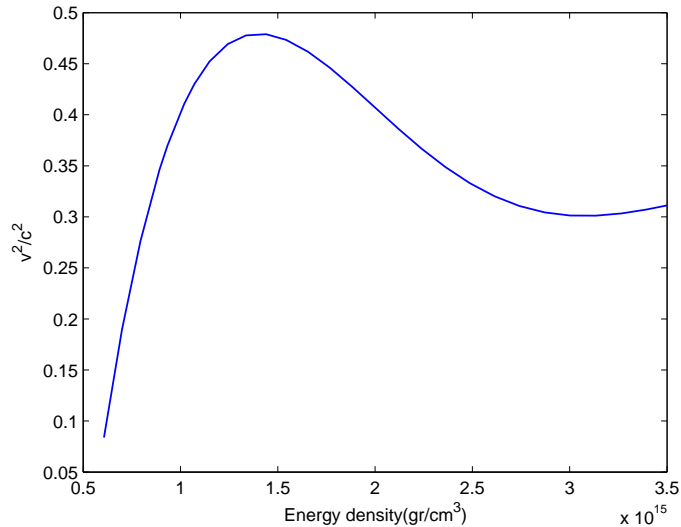}%
\end{array}
$%
\caption{Sound speed versus density.}
\label{vc}
\end{figure}

According to the stability condition for the EoS of neutron star matter in a
physically acceptable model, the corresponding extracted velocity of sound ($%
v$) must be less than the light velocity ($c$) \cite{Herrara1992, Aberu2007}%
. Thus the stability condition is given $0\leq v^{2}=(\frac{dP}{d\mathcal{E}}%
)\leq c^{2}$ or $0\leq \frac{v^{2}}{c^{2}}=\frac{1}{c^{2}}(\frac{dP}{d%
\mathcal{E}})\leq 1$. Using Eq. \ref{Prho} we compute $\frac{v^{2}}{c^{2}}$
versus density which is presented in in Fig. \ref{vc}. It is evident that
stability condition is satisfied by the EoS of neutron star with quark
matter.

\subsection{Le Chatelier's principle}

{Le Chatelier's principle is defined as: }the matter of {star satisfies $dP/d%
\mathcal{E}\geq 0$ which is a essential condition of a stable body both as a
whole and also with respect to the non-equilibrium elementary regions with
spontaneous expansion or contraction \cite{Glendenning}. According the Fig. %
\ref{vc}, the Le Chatelier's principle is established. }

The above results show that we encounter with a suitable EoS for studying
the neutron stars with a quark matter in the core. Therefore, we consider
our EoS and investigate the structure of neutron star with three different
layers.

\section{Structure of the Neutron Star with and without Quark Matter}

In this section, we study the effect of quark matter on the structure of
neutron stars, and then compare our results with the structure of neutron
stars without quark matter. Employing the obtained EoS for neutron star with
the quark matter and using TOV equation, we have gotten interesting results
which are given in Table \ref{tab2}. %
\begin{table*}[tbp]
\caption{Structure properties of neutron star without quark matter (NS) and
with quark core (NS+Q).}
\label{tab2}
\begin{center}
\begin{tabular}{||c|c|c|c|c|c|c|c|c||}
\hline\hline
Type of Star & $M_{max}\left(M_{\,\odot}\right) $ & $R\left(km\right) $ & $%
\mathcal{E}_c\left({10^{14}gr/cm^3}\right) $ & $R_{Sch}\left(km\right)$ & $%
\overline{\rho}\left({10^{14}gr/cm^3}\right)$ & $\ \ \ \sigma$ & $K(10^{-8}$
$m^{-2})$ & $z$ \\ \hline\hline
NS & 1.98 & 9.8 & 27.1 & 5.84 & 9.99 & 0.59 & 2.15 & 0.57 \\ \hline
NS+Quark Core & 1.8 & 10 & 25.8 & 5.31 & 8.55 & 0.53 & 1.84 & 0.46 \\
\hline\hline
\end{tabular}%
\end{center}
\end{table*}


\begin{figure}[tbp]
$%
\begin{array}{c}
\epsfxsize=10cm \epsffile{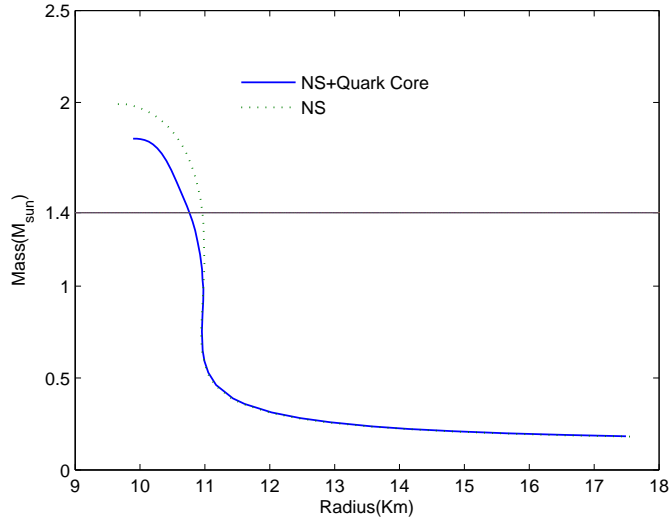}%
\end{array}
$%
\caption{Mass-radius relation for the neutron stars with and without the
quark matter.}
\label{MR}
\end{figure}
%

\begin{figure}[tbp]
$%
\begin{array}{c}
\epsfxsize=10cm \epsffile{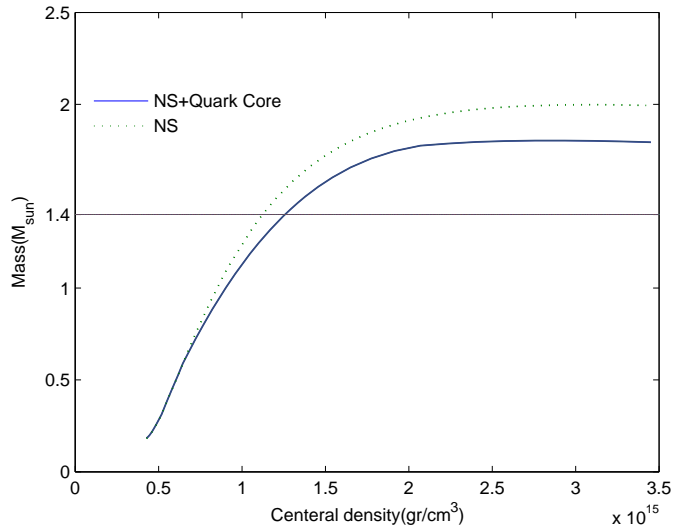}%
\end{array}
$%
\caption{Mass-the central density relation for the neutron stars with and
without the quark matter.}
\label{ME}
\end{figure}
%
%
As one can see, considering the quark matter in the core of neutron stars
change their structure properties. Indeed, there are some interesting
results when we consider the quark matter in the calculation of structure of
neutron stars. For example, by considering the quark matter, the maximum
mass decreases (see Table \ref{tab2} and Fig. \ref{MR}, for more details).
In order to do more investigation of the effect of quark matter, we
calculate another properties of the neutron star such as the average
density, compactness, Kretschmann scalar, gravitational redshift and
dynamical stability.

\subsection{Average density}

The average density of an object has the following form,
\begin{equation}
\overline{\rho }=\frac{M}{\frac{4}{3}\pi R^{3}},  \label{eq15}
\end{equation}%
where for our system, $M$ and $R$ are the gravitational mass and radius of a
neutron star. The presented results in Table \ref{tab2}, show that by
considering the quark matter in the neutron star core, the average density
from the perspective of a distant observer (or a observer outside the
neutron star) decreases. Indeed, by adding the quark matter to the core of
neutron star, the total average density (this average density is not related
to density of center of neutron star) of this system decreases.

\subsection{Compactness}

The compactness of a spherical object is usually defined as the ratio of
Schwarzschild radius to the radius of object ($\sigma =\frac{R_{Sch}}{R}$),
which may be indicated as the strength of gravity of compact objects. Our
results for the compactness are presented in Table \ref{tab2}. Similar to
the average density, by applying the quark matter in the calculation of
structure of neutron star, the compactness from the perspective of a distant
observer (or a observer outside the neutron star) decreases.

\subsection{Kretschmann scalar}

Another quantity that gives us information of the strength of gravity is
related to the spacetime curvature. According to this fact that, in the
Schwarzschild spacetime, the components of the Ricci tensor ($R_{\mu \nu }$)
and the Ricci scalar ($R_{\mu \nu }R^{\mu \nu }$) are zero outside the star,
and therefore these quantities do not give us any information about the
spacetime curvature. Therefore, we use another quantity in order to evaluate
the curvature of spacetime. The quantity that can help us to find out the
curvature of spacetime is related to the Riemann tensor ($R_{\mu \nu \gamma
\delta }$). The Riemann tensor may have more components, and also for
simplicity, we can evaluate the Kretschmann scalar for measurement of the
curvature in a vacuum. So, the curvature at the surface of a neutron star is
given as \cite{RainbowII,Psaltis,Eksi}
\begin{equation}
K=\frac{4\sqrt{3}GM}{c^{2}R^{3}}.
\end{equation}%
Our results confirm that by adding the quark matter to the core of neutron
star, the strength of gravity decreases (see Table \ref{tab2}).

\subsection{Gravitational redshift}

The gravitational redshift is given as follows,
\begin{equation}
z=\frac{1}{\sqrt{1-\frac{2GM}{c^{2}R}}}-1.
\end{equation}

The result related to the gravitational redshift are given in Table \ref%
{tab2}. This result shows that by considering a quark core inside a neutron
star, the gravitational redshift decreases.

\subsection{Dynamical Stability}

Here we use Shapiro and Teukolsky's approach \cite{Shapiro}, for studying
the dynamical stability of stellar model against the infinitesimal radial
adiabatic perturbation. In this approach, the dynamical stability condition
only require that the pressure-averaged value of the adiabatic index be more
than $4/3$, but not for the local value of the adiabatic index. This
relation is given by
\begin{equation}
\overline{\gamma }=\frac{\int_{0}^{R}\gamma pr^{2}dr}{\int_{0}^{R}pr^{2}dr},
\label{avgamma}
\end{equation}%
where it must be more than $4/3$. Here, we consider Shapiro and Teukolsky's
approach for evaluating the dynamical stability of hybrid stars extracted
from our EoS. Our calculation about the pressure-averaged value of the
adiabatic index (Eq. (\ref{avgamma})) are%
\begin{eqnarray}
\overline{\gamma }_{NS} &=&2.98, \\
&&  \notag \\
\overline{\gamma }_{NS+Q} &=&2.38.
\end{eqnarray}%
The above quantities show that the pressure-averaged value of the adiabatic
index for neutron stars with and without a quark core are $2.38$ and $2.98$,
respectively. Therefore these quantities are more than $4/3$. This shows
that the obtained pure neutron stars and hybrid stars obey the dynamical
stability.
\begin{table*}[tbp]
\caption{Properties of neutron star with the gravitational mass equal to $%
1.4M_{\,\odot }$ for NS and NS+Q.}
\label{tab3}
\begin{center}
\begin{tabular}{||c|c|c|c|c||}
\hline\hline
Type of Star & $R\left(km\right) $ & $\ \ \ \sigma$ & $K(10^{-8}$ $m^{-2})$
& $z$ \\ \hline\hline
NS & 10.96 & 0.37 & 1.09 & 0.26 \\ \hline
NS+Quark Core & 10.76 & 0.38 & 1.15 & 0.27 \\ \hline\hline
\end{tabular}%
\end{center}
\end{table*}

\begin{table*}[tbp]
\caption{Properties of neutron star with the gravitational mass equal to $%
1.8M_{\,\odot }$ for NS and NS+Q.}
\label{tab4}
\begin{center}
\begin{tabular}{||c|c|c|c|c||}
\hline\hline
Type of Star & $R\left(km\right) $ & $\ \ \ \sigma$ & $K(10^{-8}$ $m^{-2})$
& $z$ \\ \hline\hline
NS & 10.56 & 0.50 & 1.56 & 0.42 \\ \hline
NS+Quark Core & 10.00 & 0.53 & 1.84 & 0.46 \\ \hline\hline
\end{tabular}%
\end{center}
\end{table*}

Another interesting results is related to contraction of neutron stars due
to the quark core (see Tables \ref{tab3} and \ref{tab4}, for more details).
In the other words, the obtained results in Tables \ref{tab3} and \ref{tab4}
and Fig. \ref{MR} show that the radius of neutron stars without the quark
matter (NS) and with the gravitational mass equal to $1.4M_{\,\odot }$ (or $%
1.8M_{\,\odot }$) are greater than the radius of neutron stars with a quark
core (NS+Q). Indeed, for the same gravitational masses of NS and NS+Q, the
compactness, the Kretschmann scalar and the gravitational redshift increase
due to the reduced radius. In addition, this difference appears for the
neutron stars with the gravitational mass higher than the solar mass ($M\geq
M_{\,\odot }$) (see Fig. \ref{MR}).

Briefly, we can see that the existence of quark matter inside the neutron
stars leads to decreasing for the maximum mass and so it contracts them.
These results indicate that by applying the mentioned EoS in Eq. (\ref{Prho}%
), the cores of neutron stars with mass more than $2M_{\,\odot }$, do not
have any quark matter. Indeed, the presented EoS in paper (Eq. (\ref{Prho}%
)), can not predict the existence of massive hybrid stars with more than $%
2M_{\,\odot }$, because by adding the quark matter to the structure of these
stars, the maximum masses decreases.


\section{Summary and Conclusion}

The paper at hand studied the structure of cold\ hybrid stars which included
three different parts: i) a layer of hadronic matter, ii) a mixed phase of
quarks and hadrons, and iii) a quark matter in the core. For layer of
hadronic matter we used the lowest-order constrained variational (LOCV)
many-body method employing the $UV14+TNI$ potential for the nucleon-nucleon
interaction (for more details about neutron star with hadron matter and $%
UV14+TNI$ potential see ref. \cite{bordbar2006}). In another layer (mixed
phase of quarks and hadrons), we considered Gibss equilibrium condition.
Indeed for this layer, the temperature, pressures and chemical potentials of
both hadron and quark phases are equal. Finally, we considered a quark
matter in the core of neutron stars, and for this region, we used the total
energy which included up, down and strange quarks within MIT bag model. We
applied TOV equation for obtaining the structure properties of these stars.
Our results indicated that the maximum mass, the average density,
compactness, gravitational redshift, and Kretschmann scalar of neutron stars
decrease by adding a quark core to them. These results led to contraction of
hybrid stars. On the other hand, by applying the mentioned EoS in this
paper, we found an upper limit for the maximum mass of hybrid stars. In
other word, hybrid stars in our model, did not have mass more than two times
of the solar mass ($M_{\max }\leq 2M_{\,\odot }$). It is notable that, Hoyos
et al. have evaluated the properties of neutron stars by applying top-down
holographic model for strongly interacting quark matter \cite{Hoyos}. They
have obtained an EoS which was matched with state-of-the-art results for
dense nuclear. Also, they have solved\ TOV equation with their EoS, and
found the maximal stellar masses in the excess of two solar masses. Their
results showed that there are no any quark matter inside these massive
neutron stars. In other words, their results confirm our conclusions about
the existence an upper limit for neutron stars with a quark core, i.e, $%
M_{\max }\leq 2M_{\,\odot }$. However there were some difference between
their results with ours. For example, the obtained radius of neutron stars
in our calculations were greater than their results (the radius of a neutron
star with $M\simeq 2M_{\,\odot }$, was about $9.7$\ km \cite{Hoyos}, but we
found that it was about $10$\ km). Finally, we investigated the dynamical
stability of neutron stars with and without the quark core. Our calculations
indicated that these neutron stars are stable against the radial adiabatic
infinitesimal perturbations. Briefly, we obtained the quite interesting
results for hybrid stars, such as;

i) the EoS derived in this work satisfied the energy, Le Chatelier and
stability conditions.

ii) considering the EoS introduced in this paper, we found that the maximum
mass of cold hybrid stars could not be more than $2M_{\,\odot }$($M_{\max
}\leq 2M_{\,\odot }$). In other word, there are no cold massive hybrid stars
in the mass range $M_{\max }>2M_{\,\odot }$, when we applied our EoS.
Therefore, our results showed that inside the neutron stars such as $%
4U~1700-377$ \cite{ClarK} with the mass about $2.4M_{\odot }$, and $%
J1748-2021B$ \cite{J1748-2021B} with the mass about $2.7M_{\odot }$, there
is no any quark matter.

iii) the maximum mass, the average density, compactness, the Kretschmann
scalar, and gravitational redshift of neutron stars decrease owing to the
existence of quark matter in the structure them.

iv) the neutron stars are contracted due to the presence of quark matter in
their center.

v) the obtained results indicated that the studied hybrid stars in this
paper are stable against the radial adiabatic infinitesimal perturbations.

vi) for neutron stars with the gravitational mass more than one solar mass ($%
M\geq M_{\odot }$), there was a difference between NS and NS+Q. In other
words, there was no any difference between the properties of NS and NS+Q in
the range $M<M_{\odot }$.

It will be worthwhile to study the effects of the magnetic field \cite%
{MagI,MagII,MagIII,MagIV}, generalization of static compact objects to
anisotropic \cite{anisoI,anisoII,anisoIII,anisoIV,anisoV,anisoVI}, rotating
\cite{rotI,rotII,rotIII,rotIV,rotV,rotVI,rotVII,rotVIII,rotIX,rotX}, rapidly
rotating \cite%
{rapidI,rapidII,rapidIII,rapidIV,rapidV,rapidVI,rapidVII,rapidVIII} on the
structure properties of neutron stars with a quark matter in core can be
interesting topics. We leave these issues for future works.


\section*{Acknowledgements}

{We thank an anonymous referee for useful comments. We wish to thank Shiraz
University Research Council. This work has been supported financially by the
Research Institute for Astronomy and Astrophysics of Maragha (RIAAM) under
research project No. 1/5750-56. TY wishes to thank the Research Council of
Islamic Azad University, Bafgh Branch}.



\begin{thebibliography}{999}
\bibitem{Ivanenko} D. D. Ivanenko and D. F. Kurdgelaidze, Astrophys. 1
(1965) 251.

\bibitem{Itoh} N. Itoh, Progr. Theor. Phys. 44 (1970) 291.

\bibitem{Fritzsch} H. Fritzsch, M. Gell-Mann and H. Leutwyler, Phys. Lett. B
47 (1973) 365.

\bibitem{Collins} J. C. Collins and M. J. Perry, Phys. Rev. Lett. 34 (1975)
1353.

\bibitem{glendenning1992} N. K. Glendenning,\ Phys. Rev. D\textbf{\ }46
(1992) 1274.

\bibitem{Phase1} S. K. Ghosh, S. C. Phatak and P. K. Sahu, Z. Phys. A 352
(1995) 457.

\bibitem{Phase2} S. Kubis and M. Kutschera, Phys. Rev. Lett. 76 (1996) 3876.

\bibitem{Phase3} D. Bandyopadhyay, S. Chakrabarty, and S. Pal, Phys. Rev.
Lett. 79 (1997) 2176.

\bibitem{Phase4} A. Steiner, M. Prakash and J. M. Lattimer, Phys. Lett. B
486 (2000) 239.

\bibitem{Phase5} G. F. Burgio, M. Baldo, H. -J. Schulze and P. K. Sahu,
Phys. Rev. C 66 (2002) 025802.

\bibitem{Phase6} G. Lugones, T. A. S. do Carmo, A. G. Grunfeld and N. N.
Scoccola, Phys. Rev. D 81 (2010) 085012.

\bibitem{Phase7} N. Yasutake, T. Maruyama and T. Tatsumi, Phys. Rev. D 86
(2012) 101302.

\bibitem{Phase8} C. H. Lenzi and G. Lugones, Astrophys. J. 759 (2012) 57.

\bibitem{Phase9} N. Chamel, A. F. Fantina, J. M. Pearson and S. Goriely,
Proceedings of the International Astronomical Union, 8 (2012) 356.

\bibitem{Phase10} B. Franzon, R. O. Gomes and S. Schramm, Mon. Not. Roy.
Astron. Soc. 463 (2016) 571.

\bibitem{Phase11} J. P. Pereira, C. V. Flores and G. Lugones, Astrophys. J.
860 (2018) 12.

\bibitem{Phase12} E. R. Most, L. J. Papenfort, V. Dexheimer, M. Hanauske, S.
Schramm, H. Stcker and L. Rezzolla, Phys. Rev. Lett. 122 (2019) 061101.

\bibitem{Phase13} J. P. Pereira and G. Lugones, Astrophys. J. 871 (2019) 47.

\bibitem{Orsaria2013a} M. Orsaria, H. Rodrigues, F. Weber and G. A.
Contrera, Phys. Rev. D 87 (2013) 023001.

\bibitem{Orsaria2013b} M. Orsaria, H. Rodrigues, F. Weber and G. A.
Contrera, Phys. Rev. C 89 (2014) 015806.

\bibitem{Plumari} S. Plumari, G. F. Burgio, V. Greco and D. Zappala, Phys.
Rev. D 88 (2013) 083005.

\bibitem{Chen} H. Chen, J. -B. Wei, M. Baldo, G. F. Burgio and H. -J.
Schulze, Phys. Rev. D 91 (2015) 105002.

\bibitem{Lastowiecki} R. Lastowiecki, D. Blaschke, T. Fischer and T. Klahn,
Phys. Part. Nucl. 46 (2015) 843.

\bibitem{Spinella} W. M. Spinella, F. Weber, G. A. Contrera and M. G.
Orsaria, Eur. Phys. J. A 52 (2016) 61.

\bibitem{NSQMI} H. Mishra, S. P. Misra, P. K. Panda and B. K. Parida, Int.
J. Mod. Phys. E 2 (1993) 547.

\bibitem{NSQMII} M. L. Olesen and J. Madsen, Phys. Rev. D 49 (1994) 2698.

\bibitem{NSQMIII} D. Bandyopadhyay, S. Chakrabarty and S. Pal, Phys. Rev.
Lett. 79 (1997) 2176.

\bibitem{NSQMIV} M. B. Christiansen and N. K. Glendenning, Phys. Rev. C 56
(1997) 2858.

\bibitem{NSQMV} S. Pal, M. Hanauske, I. Zakout, H. Stocker and W. Greiner,
Phys. Rev. C 60 (1999) 015802.

\bibitem{NSQMVI} A. W. Steiner, M. Prakash and J. M. Lattimer, Phys. Lett. B
486 (2000) 239.

\bibitem{NSQMVII} R. Oechslin, G. Poghosyan and K. Uryu, Nucl. Phys. A 718
(2003) 706.

\bibitem{NSQMVIII} T. Harko, K. S. Cheng and P. S. Tang, Astrophys. J. 608
(2004) 945.

\bibitem{NSQMIX} J. Staff, R. Ouyed and P. Jaikumar, Astrophys. J. 645
(2006) L145.

\bibitem{NSQMIXa} A. W. Steiner, M. Prakash and J. M. Lattimer, Phys. Lett.
B 486 (2000) 239.

\bibitem{NSQMX} P. Jaikumar, S. W. Reddy and A. W. Steiner, Mod. Phys. Lett.
A 21 (2006) 1965.

\bibitem{NSQMXa} I. Bombaci, G. Lugones and I. Vidana, Astron. Astrophys.
462 (2007) 1017.

\bibitem{NSQMXI} F. Yang and H. Shen, Phys. Rev. C 77 (2008) 025801.

\bibitem{NSQMXII} M. G. Alford, Nucl. Phys. A 830 (2009) 385c.

\bibitem{NSQMXIII} G. Rupak and P. Jaikumar, Phys. Rev. C 82 (2010) 055806.

\bibitem{NSQMXIV} I. Bombaci, D. Logoteta, C. Providencia and I. Vidana,
Astron. Astrophys. 528 (2011) A71.

\bibitem{NSQMXV} G. Y. Shao, Phys. Lett. B 704 (2011) 343.

\bibitem{NSQMXVa} N. Yasutake, G. F. Burgio and H. J. Schulze, Phys. Atom.
Nucl. 74 (2011) 1502.

\bibitem{NSQMXVI} D. Logoteta, I. Bombaci, C. Providencia and I. Vidana,
Phys. Rev. D 85 (2012) 023003.

\bibitem{NSQMXVIII} G. Y. Shao, M. Colonna, M. Di Toro, Y. X. Liu and B.
Liu, Phys. Rev. D 87 (2013) 096012.

\bibitem{NSQMXXIa} B. Franzon, V. Dexheimer and S. Schramm, Mon. Not. Roy.
Astron. Soc. 456 (2015) 2937.

\bibitem{NSQMXXIb} S. M. de Carvalho, et al., Phys. Rev. C 92 (2015) 035810.

\bibitem{NSQMXXII} T. Miyatsu, M. -K. Cheoun and K. Saito, Astrophys. J. 813
(2015) 135.

\bibitem{NSQMXXIII} I. Bombaci, D. Logoteta, I. Vidana and C. Providencia,
Eur. Phys. J. A 52 (2016) 58.

\bibitem{NSQMXXIV} B. Franzon, R. O. Gomes and S. Schramm, Mon. Not. Roy.
Astron. Soc. 463 (2016) 571.

\bibitem{NSQMXXV} A. Mukherjee, S. Schramm, J. Steinheimer and V. Dexheimer,
Astron. Astrophys. 608 (2017) A110.

\bibitem{NSQMXXVI} G. Baym, T. Hatsuda, T. Kojo, P. D. Powell, Y. Song and
T. Takatsuka, Report. Progr. Phys. 81 (2018) 056902.

\bibitem{Tolman1934} R. C. Tolman, Proc. Nat. Acad. Sc. 20 (1934) 169.

\bibitem{Tolman1939} R. C. Tolman, Phys. Rev. 55 (1939) 364.

\bibitem{Oppenheimer} J. R. Oppenheimer and G. M. Volkoff, Phys. Rev.\emph{\
}55 (1939) 374.

\bibitem{TOVI} N. K. Glendenning, "\textit{Compact Star, Nuclear Physics,
Particle Physics, and General Relativity", }(Springer, New York, 2000).

\bibitem{TOVII} F. Weber, "\textit{Pulsars as Astrophysical Laboratories for
Nuclear and Particle Physics" }(Institute of Physics, Bristol, 1999).

\bibitem{TOVIII} N. Yunes and M. Visser, Int. J. Mod. Phys. A 18 (2003) 3433.

\bibitem{TOVIV} R. R. Silbar and S. Reddy, American. J. Phys. 72 (2004) 892.

\bibitem{TOVV} G. Narain, J. Schaffner-Bielich and I. N. Mishustin, Phys.
Rev. D 74 (2006) 063003.

\bibitem{TOVVII} P. Boonserm, M. Visser and S. Weinfurtner, Phys. Rev. D 76
(2007) 044024.

\bibitem{TOVIX} X. Li, F. Wang and K. S. Cheng, J. Cosmol. Astropart. Phys.
10 (2012) 031.

\bibitem{TOVXII} A. M. Oliveira, H. E. S. Velten, J. C. Fabris and I. G.
Salako, Eur. Phys. J. C 74 (2014) 3170.

\bibitem{TOVXIII} X. T. He, F. J. Fattoyev, B. A. Li and W. G. Newton, Phys.
Rev. C 91 (2015) 015810.

\bibitem{ClarK} J. S. Clark, et al., Astron. Astrophys. 392 (2002) 909.

\bibitem{J1748-2021B} P. C. C. Freire, et al., Astrophys. J. 675 (2008) 670.

\bibitem{bordbar1997} G. H. Bordbar and M. Modarres, J. Phys. G: Nucl. Part.
Phys. 23 (1997) 1631.

\bibitem{bordbar1998} G. H. Bordbar and M. Modarres,\ Phys. Rev. C\textbf{\ }%
57 (1998) 714.

\bibitem{modarres1998} M. Modarres and G. H. Bordbar, Phys. Rev. C\textbf{\ }%
58 (1998) 2781.

\bibitem{bigdeli2009} M. Bigdeli, G. H. Bordbar and Z. Rezaei,\emph{\ }Phys.
Rev. C 80 (2009) 034310.

\bibitem{clark} J. W. Clark, Prog. Part. Nucl. Phys. 2 (1979) 89.

\bibitem{bordbar2006} G. H. Bordbar and M. Hayati, Int. J. Mod. Phys. A 21
(2006) 1555.

\bibitem{chodos1974} A. Chodos, R. L. Jaffe, K. Johnson, C. B .Thorn and V.
F. Weisskopf, Phys. Rev. D\textbf{\ }9 (1974) 3471.

\bibitem{baym1985} G. Baym, E. W. Kolb, L. McLerran, T. P. Walker and R. L.
Jaffe,\emph{\ }Phys. Lett. B\textbf{\ }160 (1985) 181.

\bibitem{TOVVI} G. H. Bordbar, M. Bigdeli and T. Yazdizade, Int. J. Mod.
Phys. A 21 (2006) 5991.

\bibitem{burgio2002} G. F. Burgio, M. Baldo, P. K. Sahu and H. J. Schulze,
Phys. Rev. C\textbf{\ }66 (2002) 025802.

\bibitem{Poisson} E. Poisson, "\textit{A relativist`s toolkit}", (Cambrige
university press , New York, 2004).

\bibitem{Herrara1992} L. Herrera, Phys. Lett. A{\textbf{\ }165} (1992) 206.

\bibitem{Aberu2007} H. Abreu, H. Hernandez and L. A. Nunez, Class. Quantum
Grav. 24{\ (2007)} 4631.

\bibitem{Glendenning} N. K. Glendenning, Phys. Rev. Lett. 85 (2000) 1150.

\bibitem{RainbowII} B. Eslam Panah, et al., Astrophys. J. 848 (2017) 24.

\bibitem{Psaltis} D. Psaltis, Liv. Rev. Relativ. 11 (2008) 9.

\bibitem{Eksi} K. Y. Eksi, C. Gungor and M. M. Turkoglu, Phys. Rev. D 89
(2014) 063003.

\bibitem{Shapiro} S. L. Shapiro and S. A. Teukolsky, "\textit{Black Holes,
White Dwarfs and Neutron Stars: The Physics of Compact Objects}", Wiley, New
York (1983).

\bibitem{Hoyos} C. Hoyos, D. R. Fernandez, N. Jokela and A. Vuorinen, Phys.
Rev. Lett. 117 (2016) 032501.

\bibitem{MagI} M. Bocquet, S. Bonazzola, E. Gourgoulhon and J. Novak,
Astron. Astrophys. 301 (1995) 757.

\bibitem{MagII} J. S. Heyl and S. Kulkarni, Astrophys. J. Lett. 506 (1998)
L61.

\bibitem{MagIII} C. Y. Cardall, M. Prakash and J. M. Lattimer, Astrophys. J.
554 (2001) 322.

\bibitem{MagIV} D. Chatterjee, T. Elghozi, J. Novak and M. Oertel, Mon. Not.
Roy. Astron. Soc. 447 (2015) 3785.

\bibitem{anisoI} J. A. Miralles, J. A. Pons and V. Urpin, Astron. Astrophys.
420 (2004) 245.

\bibitem{anisoII} C. G. Boehmer and T. Harko, Class. Quantum Grav. 23 (2006)
6479.

\bibitem{anisoIII} B. C. Paul and R. Deb, Astrophys. Space Sci. 354 (2014)
421.

\bibitem{anisoIV} S. K. Maurya, Y. K. Gupta, S. Ray and B. Dayanandan, Eur.
Phys. J. C 75 (2015) 225.

\bibitem{anisoV} D. K. Matondo and S. D. Maharaj, Astrophys. Space Sci. 361
(2016) 221.

\bibitem{anisoVI} G. Estevez-Delgado and J. Estevez-Delgado, Eur. Phys. J. C
78 (2018) 673.

\bibitem{rotI} H. Heiselberg and M. Hjorth-Jensen, Phys. Rev. Lett. 80
(1998) 5485.

\bibitem{rotII} E. Chubarian, H. Grigorian, G. S. Poghosyan and D. Blaschke,
Astron. Astrophys. 357 (2000) 968.

\bibitem{rotIII} N. Andersson and G. L. Comer, Class. Quantum Grav. 18
(2001) 969.

\bibitem{rotIV} Z. B. Etienne, Y. T. Liu and S. L. Shapiro, Phys. Rev. D 74
(2006) 044030.

\bibitem{rotV} J. Zdunik, M. Bejger, P. Haensel and E. Gourgoulhon, Astron.
Astrophys. 450 (2006) 747.

\bibitem{rotVI} P. Pani and E. Berti, Phys. Rev. D 90 (2014) 024025.

\bibitem{rotVII} R. F. P. Mendes, G. E. A. Matsas and D. A. T. Vanzella,
Phys. Rev. D 90 (2014) 044053.

\bibitem{rotVIII} A. Cisterna, T. Delsate, L. Ducobu and M. Rinaldi, Phys.
Rev. D 93 (2016) 084046.

\bibitem{rotIX} J. G. Coelho, et al., Astron. Astrophys. 599 (2017) A87.

\bibitem{rotX} K. V. Staykov, D. Popchev, D. D. Doneva and S. S. Yazadjiev,
Eur. Phys. J. C 78 (2018) 586.

\bibitem{rapidI} D. Lai and S. L. Shapiro, Astrophys. J. 442 (1995) 259.

\bibitem{rapidII} S. Yoshida, S. Karino, S. Yoshida and Y. Eriguchi, Mon.
Not. Roy. Astron. Soc. 316 (2000) L1.

\bibitem{rapidIII} J. Zdunik, P. Haensel, E. Gourgoulhon and M. Bejger,
Astron. Astrophys. 416 (2004) 1013.

\bibitem{rapidIV} J. Zdunik, M. Bejger, P. Haensel and E. Gourgoulhon,
Astron. Astrophys. 479 (2008) 515.

\bibitem{rapidV} P. Haensel, J. Zdunik, M. Bejger and J. Lattimer, Astron.
Astrophys. 502 (2009) 605.

\bibitem{rapidVI} D. D. Doneva, S. S. Yazadjiev, N. Stergioulas and K. D.
Kokkotas, Phys. Rev. D 88 (2013) 084060.

\bibitem{rapidVII} B. Kleihaus, J. Kunz, S. Mojica and M. Zagermann, Phys.
Rev. D 93 (2016) 064077.

\bibitem{rapidVIII} S. -S. Luk and L. -M. Lin, Astrophys. J. 861 (2018) 141.
\end{thebibliography}
\end{document}